\documentclass[preprint,aps,a4paper]{revtex4}

\usepackage[dvips]{graphics,epsfig}

\begin{document}

\title{The two envelopes probability paradox: Much ado about nothing}
\author{R.A. V\'azquez$^1$} \affiliation{Departamento de F\'{i}sica de
  Part\'{i}culas and\\ 
  Instituto Galego de F\'{\i}sica de Altas Enerx\'{\i}as, \\ 
  Campus Sur\\ 
  15782 Santiago de Compostela, Spain\\ 
{\small $^1$vazquez@fpaxp1.usc.es} }

\begin{abstract}
The two envelopes paradox is discussed. By calculating the conditional
probability, we arrive at a conditional expectations which differs from 
existing results. 
\end{abstract}

\maketitle

\section{Introduction}

The two envelopes paradox is a famous problem of probability which has been
extensively discussed in the literature, see for instance
\cite{Chalmers,Utts,Broome}. Here it is presented as a pedagogical
exercise. Although the statement of the problem is simple the results are
rather counterintuitive and paradoxical at first: an extremely careful
treatment of the problem is needed to avoid the pitfalls where many experts on
the field have fallen. It can be used, therefore, as a sharpener of the
student intuition on probability theory. Too often statistics and probability
theory are taught in the physics degree as a cookbook of recipes where one
result has nothing to do with the next.

Most of the presumed paradoxical results for this problem are related to uses
and abuses of (i) the conditional probability and conditional expectation (ii)
subtleties associated with non normalizable distributions or distributions
with no moments (iii) the bad use of Bayes theorem and of prior
probabilities. In addition, we will show that results existing in the
literature for the uniform and the scale invariant prior probabilities are
incorrect, due to the lack of a proper normalization of integrals. 

We will start by stating the paradox in the two most common situations:
\begin{center}
{\it (S0)} A person tells you the following: She will think of a number $x$
and write down two checks. In one she writes down the amount of $x$ dollars,
in the other she writes down the amount of $2 x $ dollars. The checks are put
into two envelopes and closed. You can pick out any of the envelopes and keep
the amount of that check. Once you have decided and chosen one of the two
envelopes, but before opening it, she tells you whether you prefer to change
your decision and pick the other envelope.  Your goal is to maximize the
amount of money you can win.  What should you do?
\end{center}

This statement is often complemented with the following
\begin{center}
{\it (S1)} The same as above. But you are allowed to open your first picked
envelope and read the amount written on the check, say $y$ dollars. Again, you
are asked whether you prefer to switch to the other envelope or stick with the
first one.
\end{center}

The paradox in {\it (S1)} arises as the standard reasoning goes:
Since I do not know $x$, I assume an uniform probability for $x$. The check I
have chosen (and has a value of $y$, known to me) has a probability 1/2 of
being $x$ and a probability 1/2 of being $2 x$. So the other check has a
probability 1/2 of being $y/2$ and a probability 1/2 of being $2
y$. Therefore, the expectation (the average value) of the other envelope (say
$z$)  is
given by:
\begin{equation}
<z>_y = \frac{1}{2} \; ( 2 y +  y/2 ) = \frac{5}{4} y \; > y.
\label{wuniform}
\end{equation}
Here, I have added a subscript $y$ to the average symbol to clearly state that
the expectation is taken with $y$ constant (and known).  Therefore,
independently of the value of $y$, the {\it other} envelope has an expectation
value greater than $y$. This is the paradoxical situation.  The paradox goes
on by saying that since the result above gives me an expectation greater than
$y$ for any value of $y$, then I do not need to open the envelope at all, and
therefore in the statement {\it (S0)} I should have $<z> = 5/4 y$, whether or
not I have opened the envelope. But this is really paradoxical since nothing
tells me to prefer the first envelope in the first place. I could have chosen
the second envelope first and the reasoning would be the same: no matter which
envelope I have chosen {\it the other} appears to have a larger expectation
value.

The above arguments are twofold wrong. First, the statement {\it (S0)} is
completely symmetrical in $y$ and $z$. Any calculation that give different
values to the expectation of $y$ and $z$ is not justified. And second, I will
argue that the calculation of the expectation of $z$ for a given $y$ is
incorrect. In particular, a uniform distribution does not produce a
conditional expectation $<z>_y = 5/4 y$. Even such authoritative sources as
\cite{Streater} give this incorrect result.

The paradox is often enhanced by the introduction of improper (non
normalizable) prior probabilities, probabilities with infinite mean, and the
incorrect treatment of conditional probabilities. 

\section{Generalities}

To fix notation, I will always call $x$ the number chosen in the first place. 
$y$ will denote the value written in the first check (the check of my
choice), and $z$ will denote the value in the other check. Evidently, either
$y = x$ and $z = 2 x$ or $y = 2 x$ and $z = x$.  I will call $f(x)$ the prior
probability of $x$. We will evaluate the expected values of $z$ and $y$ for
different choices of the prior probability. \footnote{For those readers that
  do not accept prior probabilities at all, $x$ will be considered as a
  random number with a distribution given by $f(x)$. }

Then, the probability of obtaining the three numbers $x$, $y$, $z$, is given
by
\begin{equation}
P(x,y,z|I) = \frac{f(x)}{2} \left( \delta(y-x)  \delta(z-2x) + \delta(y-2 x) 
\delta(z-x)  \right).
\label{prob}
\end{equation}
Here $I$ is the information we have about the problem such as that given by
the statement {\it (S0)}, complemented by the knowledge of the prior
probability, etc. We use the conventional notation for conditional
probabilities, $P(A|B) $ reads as the probability of $A$ given that $B$ is
true.  The two conditions (either one of the envelope has $x$ and the other
$2x$ ) have been included in the equation as a combination of Dirac deltas,
although one could very well dispose of the use of Dirac deltas, it simplifies
the notation putting all the cases into a single line. 

All the problem consists on calculating marginal probabilities from
the Eq.(\ref{prob}). Notice that the above probability is correctly
normalized, if $f(x)$ is
\begin{equation}
\int dx dy dz \; \; P(x,y,z|I) = \int dx \; f(x) = 1. 
\label{intprob}
\end{equation}

For any given prior distribution we can calculate the expectation of either
$y$ or $z$ by integrating Eq.(\ref{prob}), for instance
\begin{equation}
<z> = \int dx dy dz \; \; z \; P(x,y,z|I) = \int dx dz  \; z \; 
\frac{f(x)}{2} \; 
\left(\delta(z-2x) + \delta(z-x)  \right) = \frac{3}{2} <x>.
\label{avez}
\end{equation}
The same result is obtained for $y$ so that, $<z>= <y>$ independently of the
prior distribution $f(x)$. This answers statement {\it (S0)}: if the envelope
is not open switching envelopes has no effect on the expectation value. The
problem is completely symmetrical with respect to the two envelopes. Using 
an improper prior does not invalidate this result.

We can integrate out over $x$ in Eq.(\ref{prob}) to give
\begin{equation}
P(y,z|I) = \int dx \; P(x,y,z|I) = \frac{1}{2} f(y) \; \delta(z- 2y) 
+ \frac{1}{2} f(z) \; \delta(y- 2z). 
\label{yzmargprob}
\end{equation}
By further integrating with respect to $z$ one obtains
\begin{equation}
P(y|I) = \int dz P(y,z|I) = \frac{1}{2} f(y)  + \frac{1}{4} f(y/2).
\label{ymargprob}
\end{equation}
This is the marginal probability of $y$.  The $1/4$ is at first surprising,
but correct. It is the responsible for the incorrectness of
Eq.(\ref{wuniform}). The additional 1/2 which appears in Eq.(\ref{ymargprob})
is due to the integration of the $\delta(y -2 z)$ which indeed produce an
additional factor of 1/2. Notice that Eq.(\ref{ymargprob}) is correctly
normalized to one, since $f(y)$ is a probability density, by changing the
scale of the variable one needs to rescale also the density.

With this result we can obtain the conditional probability of $z$ given $y$,
which is the relevant probability for problem {\it (S1)}, using Bayes' theorem
\begin{equation}
P(z| y I) = \frac{P(y,z|I)}{P(y|I)}.
\label{bayes}
\end{equation}
$P(y|I)$ acts here as a normalizing factor. The expectation of $z$ given that 
$y$ is known is given by
\begin{equation}
<z>_y = \int dz \; z \; P(z|y I) = y \; \frac{1}{P(y|I)}  \; \left( f(y) 
+ \frac{1}{8} f(y/2) \right).
\label{zcond}
\end{equation}
Obviously, in general, we have $<z>_y \neq y$. So the paradox seems to be
alive. 

\section{A finite case}

As a warm up exercise consider the case where $f(x)$ is a well behaved
distribution and has finite moments. For the sake of definiteness we will
chose a simple exponential
\begin{equation}
f_0(x) = \lambda e^{- \lambda x}.
\label{finite}
\end{equation}
We assume that $\lambda$ is known to us. By integrating the above equations we
obtain
\begin{equation}
P(y| I_0) = \frac{1}{2} \; \lambda e^{-\lambda y} + \frac{1}{4} \; 
\lambda e^{-\lambda y/2}.
\label{Pfinite}
\end{equation}
Finally we get from Eq.(\ref{zcond}), after some algebra
\begin{equation}
<z>_y = \frac{y}{2} \; \; \frac{1 + 8 e^{-\lambda y/2}}{1+ 2 e^{-\lambda y/2}}.
\label{zf0}
\end{equation}
As expected from the previous section $<z>_y \neq y$. Are we getting a
paradoxical result? Of course not!  The distribution $f_0(x)$ has an average
value of $1/\lambda$ and the peak probability occurs for $x=0$, so knowing the
value of $y$ is very relevant to the inference of possible $x$'s . If $y$ is
small ($\lambda y/2 \ll 1 $ ) then it is more probable than $x = y$ than the
other way around, and switching is the best course of action. Indeed from
Eq.(\ref{zf0}) for $y$ small we have $<z>_y \sim 3/2 y $. On the other hand,
larger values of $x$ are exponentially suppressed, so that if $y$ is large
($\lambda y/2 \gg 1 $ ) it is more probable than $x$ is the lowest value of
the two allowed possibilities ( i.e.  $x = y/2$) and therefore switching is
out of the question.  Indeed, in this case $<z>_y \sim y/2 $ for $y$ large.
For a frequentists approach, a single realization of the experiment is
irrelevant. If we were to repeat this game many times we will have to average
over the marginal probability of getting a $y$
\begin{equation}
<z> = \int dy \; <z>_y P(y|I_0) \; = \; <y>,
\label{zf1}
\end{equation}
which of course reduces to our original expectation: If we repeat the
experiment many times, always switching from our first choice, the expectation
of our profit is completely symmetrical and we win nothing, on average, with
respect to not switching.

\section{The uniform prior}

Consider the case where the prior probability of getting the value $x$ is
known and uniform $f(x) = {\rm constant}$. Since the uniform distribution is
not normalizable we will choose a maximum allowed value $L$ to make a
normalizable distribution, with the hope that, after our calculation is done,
we can safely take the limit $L \rightarrow \infty$. Set
\begin{eqnarray}
f_u(x)  = &  1/L,   \; \; \; \; x<L\\ \nonumber
       =  &  0,   \; \; \; \; x >L. 
\label{uniform}
\end{eqnarray}
Integrating the expression Eq.(\ref{prob}) over $x$ and $z$ we obtain the
marginal probability for $y$
\begin{eqnarray}
P(y| I_u) = &  \frac{3}{4L},  \; \; \; \; 0<y<L\\ \nonumber
          = &  \frac{1}{4L},  \; \; \; \; L<y<2L . 
\label{yuniform}
\end{eqnarray}
Observe that the above equation is correctly normalized
and that the average value of $y$ is indeed $3/4 L = 3/2 <x>$. 
As before, we can calculate the conditional probability of the other choice
knowing $y$
\begin{eqnarray}
P(z| y I_u) =  & \frac{1}{2 L \, p(y|I_u) } \left( \frac{1}{2} 
\delta(y -z/2) + \delta(y -2 z) \right), \; \; \; \; 0< z< L \\ \nonumber
            =  & \frac{1}{2 L \, p(y|I_u) } \left( \frac{1}{2} \delta(y
-z/2) \right), \; \; \; \;  L< z< 2L. \\ \nonumber
\label{Pzuni}
\end{eqnarray}
Notice that for $z>L$ there is only one possibility. Indeed if we know that, 
$z$ must be the $2x$ and $y$ must be $x$. 
The conditional expectation is now given by
\begin{eqnarray}
<z>_y   = & \frac{3}{2} y, \; \; \; \; 0< y < L \\ \nonumber
        = & \frac{1}{4} y,   \; \; \; \; L < y < 2L.   \\ \nonumber
\label{zuni}
\end{eqnarray}
Our claim is that this equation is the correct result for an uniform prior
probability, instead of the common result Eq.(\ref{wuniform}). We can convince
ourselves by noting that all the probabilities, total and marginal are
correctly normalized and the limits of integration have been carefully taken
into account. Failure to do so leads to the incorrect result
Eq.(\ref{wuniform}). The major source of confusion comes from the failure to
recognize that if $f(x)$ is a correctly normalized probability distribution,
then $f( ax)$ is not (for $a\neq 1$). There is a Jacobian involved which our
use of delta functions have taken into account.  Additionally, one can check
that, by averaging over $y$, with the marginal probability
Eq.(\ref{yuniform}), one obtains $<z>= <y>= 3/2 <x>$ for any $L$, as it
should.

Again we see here the same behavior that in the previous example. There is a
``safe'' region (here $y <L $) where switching is profitable, however there is
also an ``unsafe'' region ($y >L$) where switching would lead to lesser your
expectations.  The gain and losses of the two regions conspire so that at the
end, if we were to repeat the game many times, we would get $<y>= <z>$. 

After all this labor, we realize that setting $L = \infty$ will not work. One
would miss the ``unsafe'' region, needed to satisfy the normalization of the
probability and the symmetry argument. One can check that the total
probability that goes into the ``unsafe'' region is constant (1/4) and
independent of $L$. So by passing to the limit one would, literally, put this
region under the rug. The fact that an uniform distribution extending to
infinity can not be normalized has precluded previous work from recognizing
it.

\section{The Jeffreys prior}

\begin{figure}
\begin{center}
\mbox{\epsfig{figure=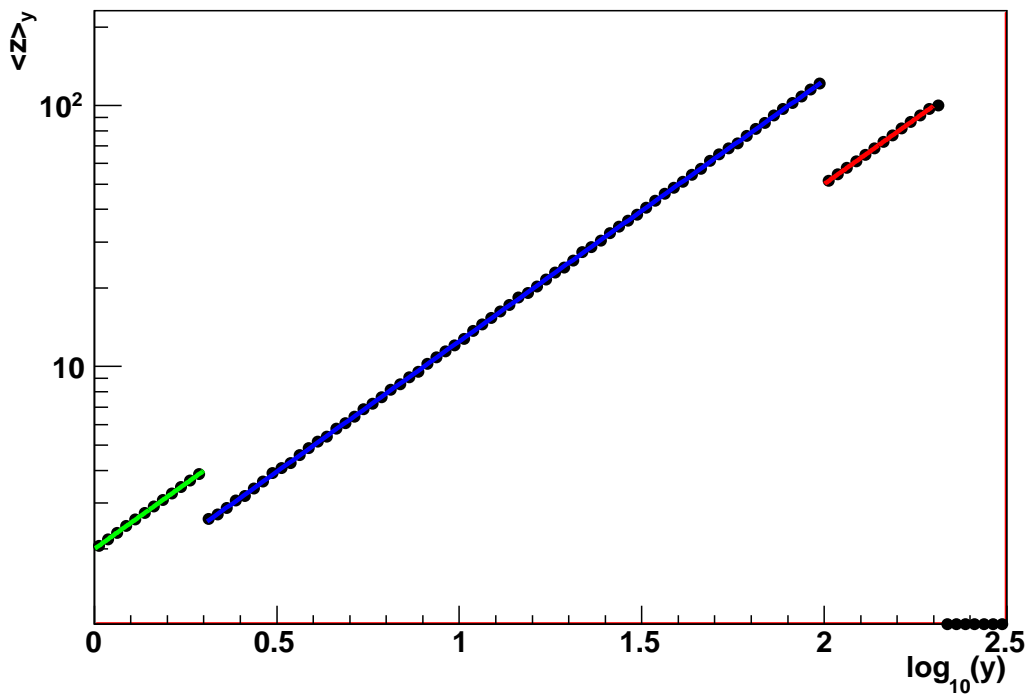,width=15.0cm}}
\label{fig1}
\caption{Results of a Monte Carlo run with $10^6$ realizations and using the
  Jeffreys prior with cut off $L_0=1$ and $L_1=100$. Shown is the value of
  $<z>_y$ as a function of $log_{10}(y)$ (dots). The lines are the result of
  Eq.(\ref{zJeff}) for the three ranges indicated: $L_0 < y < 2 L_0$ (Green) 
$2 L_0 < y< L_1$ (Blue), and $ L_1 < y < 2 L_1$ (Red). }
\end{center}
\end{figure}

In problems where scale invariance is an issue, it is convenient to introduce
the Jeffreys' prior $f(x) \propto 1/x$. The use of this probability density is
discussed in \cite{Jeffreys,Jaynes} where several examples of usage and
properties are included. It is invariant against changes of variables of the
form $x'= x^a$, and also for changes of scale $x' = b x$. In our problem,
there is no reference to any scale, $x$ could be millions or could be a penny,
so the choice of the Jeffreys prior seems appropriate. Unfortunately, it is
not normalizable and the probability diverges both for $x$ very large and for
very small values. So, we will start with a conveniently normalized prior by
setting a minimum $L_0$ and maximum $L_1$ amount of money
\begin{eqnarray} \nonumber
f_J(x)  = &  0 ,  \; \; \; \; 0 < x<L_0, \\
      = &  B/x,   \; \; \; \; L_0 < x< L_1, \\ \nonumber
       = &  0,  \; \; \; \; x >L_1. \\ \nonumber
\label{PJeffreys}
\end{eqnarray}
$B$ is the normalization constant. 
Repeating the previous steps we get
\begin{eqnarray} \nonumber
P(y|I_J) = &  B/(2y),  \; \; \; \;L_0 < y< 2 L_0, \\
      = &  B/y,   \; \; \; \; 2 L_0 < y< L_1, \\ \nonumber
      = & B/(2y),  \; \; \; \; L_1 < y < 2 L_1. \\ \nonumber
\label{yPJeffreys}
\end{eqnarray}
$P$ is zero outside the given ranges. For the conditional probability
\begin{eqnarray} \nonumber
P(z|y I_J) = &  \delta(z- 2 y) ,  \; \; \; \; L_0 < z < 2 L_0, \\
      = & \frac{1}{2} (\delta(z-2 y)+ \delta(z-y/2)),   \; \; \; \; 
 2 L_0 < z< L_1, \\ \nonumber
     =  &  \delta(z-y/2),  \; \; \; \; L_1 < z < 2 L_1. \\ \nonumber
\label{zPJeffreys}
\end{eqnarray}
And finally
\begin{eqnarray} \nonumber
\label{zJeff}
<z>_y & = 2 y ,  \; \; \; \; L_0 < y < 2 L_0, \\
      & = 5/4 y ,   \; \; \; \; 2 L_0 < y< L_1, \\ \nonumber
      & = y/2 ,  \; \; \; \; L_1 < y < 2 L_1. \\ \nonumber
\end{eqnarray}
As before the meaning of this equation is clear. If $y$ is low ($y < 2 L_0$),
the best option would be to switch, since we must have $y = x$. If $y$ is
large ($y > L_1$) then it is sure that $z=x$ and therefore switching is not
desirable. For intermediate values one gets that, on average, switching is the
preferred option. In fig.(\ref{fig1}) we show the result of a Monte Carlo run
for the Jeffreys prior with $L_0=1$ and $L_1=100$. The lines are not fits to
the result of Eq.(\ref{zJeff}).

Again the passage to the limit $L_1 \rightarrow \infty$ and $L_0 \rightarrow
0$ is not possible, but for any finite values of the limits, we get that, as
before
\begin{equation}
<z>= \int dy <z>_y \; P(y|I_J) = <y> = 3/2 <x>.
\label{zJtot}
\end{equation}
It has become a standard lore that the Jeffreys prior gives an even
expectation (i.e. $<z>_y = y$) \cite{Streater,Chalmers,Utts}. Here we have
shown that this is incorrect, those calculations miss the Jacobian in the term
having $f(2y)$. In addition, it is rather unfortunate that the ``standard''
result for the uniform prior Eq.(\ref{wuniform}) which is also wrong,
reproduces the correct result for the Jeffreys prior, in the ``safe'' region
(Eq.(\ref{zJeff})).

We could ask whether any prior which gives the ``naively expected'' result
$<z>_y = y$ exists. By direct inspection of Eq.(\ref{zcond}) we obtain that
$f_I(x) \propto 1/x^2$ would produce the desired result in the ``safe''
region. Although this distribution is not normalizable we can take the limit
so that one keeps $<z>_y = y$ and $<z>= <y>$. However, I can not think of any
physical argument that would lead us to this prior probability.

\section{Another twist}

In the computer age, we could change this game of envelopes and
checks by a computer program. But now we are allowed one further subtlety
\begin{center}
{\it (S2)} Someone chooses a number $x$.  Two options (1) or (2) (one with $x$
and the other with $2x$, as before) are presented to you in a computer
terminal. Once you choose one option and click into it, say option (1), the
computer changes the value associated to options (1) and (2) as follows. Your
choice is always set to \$10, and shown to you.  The other (hidden) option is
set either to \$20 or to \$5 depending on whether that option had
originally being set with $2x$ or $x$ respectively.  You are informed of the
whole procedure and again you are asked whether you prefer to change you
choice or not.
\end{center}
One must recognize that the game played here is different from the one in {\it
  (S1)}. In fact by changing you election from $y$ to \$10 one nullifies the
original choice of $x$ and any prior probability we may choose for it is
irrelevant. An elementary calculation shows that now the expected value of the
other choice is given by
\begin{equation}
<z>_{y=10}= \frac{1}{2} \;  2 y +\frac{1}{2} \; y/2 = \frac{5}{4} y = \; \$
\; 12.5. %
\label{twisted}
\end{equation}
It is interesting to recognize in this equation the original (and incorrect)
result Eq.(\ref{wuniform}), for an uniform prior for problem {\it
(S1)}. Apparently the reasoning behind it is something similar to this: ``You
have chosen some value $x$, since I assume the probability of a given $x$ to
be uniform, I can safely set it to anything I like, so, I open my envelope and
find, say, \$10 on it the only options is for the other choice to be \$20 or
\$5 with probabilities of 50\% each...``. As was shown above this kind of
reasoning is incorrect.

Now $<z>_y$ is greater than $y$ for any $y$ and therefore we are back to the
paradox, so it seems.  No matter what choice I made, I would be better off by
switching.  But of course, this is the correct, and expected, result: the
problem has lost the symmetry between the two choices. After all the computer
is tricking us and in fact it is putting more money, on the average, on the
{\it other} choice.

The procedure of changing whatever value of $y$ to \$10 reminds us of a
normalization or of a change of scale. The only scale invariant prior
probability is given by Jeffreys' prior. It is reassuring that we obtain here
the same result as with Jeffreys Eq.(\ref{zJeff}) in the ``safe'' region. This
could be expected. However, for any prior probability we set in the {\it
  (S1) } game there must exists also an ``unsafe'' region, where switching
will lead us to a loss, to recover the symmetry.  In short, the game played in
statement {\it (S2)} can not be cast into a {\it (S1)} game for any choice of
prior probabilities. The closest we can get is by using the scale invariant
Jeffreys prior. It is interesting to notice that intuitively one transforms an
{\it (S1)} problem with an uniform prior into a scale invariant {\it (S2)}
problem in a rather unconsciously way. This solves the paradox.

\section{Discussion}

Improper priors have been severely criticized by \cite{Streater} on
the basis that they are meaningless, since they are not normalizable.  This is
taken as a proof that the Bayesian interpretation of probabilities is
incorrect. However, improper priors can be useful and give meaningful
results. The fact that in this problem we get something contrary to our
uneducated intuition does not invalidate the use of priors, normalizable or
not. Rather, it invalidates our intuition.

Arguably, the most used wave function in quantum mechanics is the plane wave,
which is an uniform distribution over the whole space. The fact that it is not
normalizable does not deter to a physicist to calculate the cross section of,
say, an $e^+ e^-$ collision at CERN, without worrying much about the
probability of finding the electron near the Andromeda galaxy. Or again, a
charged particle traversing matter suffers from loss of energy by the emission
of radiation, the so called bremsstrahlung radiation. The probability of
emission of a photon of frequency $\omega$ is proportional to $1/\omega$, which
is of the Jeffreys' type. Here the point is that despite the infinities that
may appear (infinity number of photons emitted at very low frequency!) one can
calculate the total energy lost, which is well defined and finite (there is a
natural high frequency cut off because the total available energy is finite).
The same kind of reasoning goes behind the calculation of probabilites using
the anthropic principle or the parton distribution functions. They have
associated probabilities densities which, in some cases, can be improper. 

The moral of this is that an improper probability distribution can be safely
used by starting with a normalized distribution, doing the calculation and
then passing to the limit. If the passage to the limit gives a sensible
result, its use is justified.  In our problem, we have learnt that any hint on
the existence of a ``scale'' will give us an advantage over the indifferent
choice. For instance, we may be playing a real game with real money involved,
then there is a maximum value $x$ can have. No one will sign a check with,
say, the US GDP. Or we may be in a TV show, we know the prize history of the
show and we know that this show has never given prizes for more than, say
\$1000. Any hint will let you take advantage and win, on average. However, if
we have no idea of any scale of the money involved then the problem is not
well defined: we have no rational basis for any preferred choice. For a scale
free problem, the Jeffreys prior should be used, however, as we have seen, it
does not give a sensible result in the passage to the infinity limit, since
the expectations are infinity. If I were to set the following problem: I think
of a number $x$, estimate what is the expectation value of $x$? There is no
prior which would produce any sensible result. This is not a fault on the use
of improper priors but rather on the ill definition of our problem. 

The paradox arises because one, consciously or unconsciously, changes from an
{\it (S1)} type problem to an {\it (S2)} problem, making it asymmetric on the
two envelopes. A correct treatment of the prior probabilities reveals the
falsity of the argument. Therefore, in the absence of any additional scale,
any argument leading to a preferable choice can not be justified. 

{\bf Acknowledgments} I thank J. Sanchez-Guillen for discussions on the
problem and for reading the manuscript.


\end{document}